\documentclass[12pt]{spieman}  
\usepackage{amsmath,amsfonts,amssymb}
\usepackage{graphicx}
\usepackage{setspace}
\usepackage{tocloft}
\usepackage{subcaption}
\usepackage{siunitx}
\usepackage{placeins}
\usepackage{orcidlink}

\title{Measurement and simulation of charge diffusion in a small-pixel charge-coupled device}

\author[a,*]{Beverly J. LaMarr}
\author[a]{Gregory Y. Prigozhin}
\author[a]{Eric D. Miller\,\orcidlink{0000-0002-3031-2326}}
\author[b]{Carolyn Thayer}
\author[a]{Marshall W. Bautz\,\orcidlink{0000-0002-1379-4482}}
\author[a]{Richard Foster\,\orcidlink{0000-0002-2731-9295}}
\author[a]{Catherine E. Grant\,\orcidlink{0000-0002-4737-1373}}
\author[a]{Andrew Malonis}
\author[c,$\dagger$]{Barry E. Burke}
\author[c]{Michael Cooper}
\author[c]{Kevan Donlon}
\author[c]{Christopher Leitz}

\affil[a]{Kavli Institute for Astrophysics and Space Research, Massachusetts Institute of Technology, 77 Massachusetts Ave, Cambridge, MA, USA, 02139}
\affil[b]{Department of Climate and Space Sciences and Engineering, University of Michigan, 2455 Hayward St., Ann Arbor, MI USA 48109  }
\affil[c]{Lincoln Laboratory, Massachusetts Institute of Technology, 244 Wood St., Lexington, MA USA 02421}

\cftpagenumbersoff{figure}
\cftpagenumbersoff{table}

\newcommand{\mic}[1]{\SI{#1}{\um}}
\newcommand{\vs}{V$_{\rm{sub}}$}

\begin{document} 
\maketitle

\begin{abstract}
Future high-resolution imaging X-ray observatories may require detectors with both fine spatial resolution and high quantum efficiency at relatively high X-ray energies (E $\ge 5 $ keV ). A silicon imaging detector meeting these requirements will have a ratio of detector thickness to pixel size of six  or more, roughly twice that of legacy imaging sensors. The larger aspect ratio of  such a sensor's detection volume implies greater diffusion of X-ray-produced charge packets.  We investigate consequences of  this fact for sensor performance, reporting charge diffusion measurements in a fully-depleted back-illuminated CCD with a thickness of \mic{50} and  pixel size of \mic{8}. We are able to measure the size distributions of charge packets produced by 5.9 keV and 1.25 keV X-rays in this device. We find that individual charge packets exhibit a gaussian spatial distribution, and determine the frequency distribution of event widths for a range of detector bias (and  thus internal electric field strength) levels. At the largest bias, we find a  standard deviation for the largest charge packets (produced by X-ray interactions closest to the entrance surface of the device) of \mic{3.9}. We show that the shape of the event width distribution provides a clear indicator of full depletion, and use a previously developed technique to infer the relationship between event width and interaction depth.  We compare measured width distributions to simulations. Although  we can obtain  good agreement  for a given detector bias, with our current simulation we are  unable to fit the data for the full range of of bias levels  with a single set of simulation parameters. We compare traditional, 'sum-above-threshold'  algorithms for individual event amplitude determination to  gaussian fitting of individual events and  find that better  spectroscopic  performance is  obtained with the former for 5.9 keV events, while the two methods provide comparable results  at 1.25 keV. The reasons for this difference are discussed. We point out the importance of read-noise driven charge detection thresholds in degrading spectral resolution , and note that the derived read noise requirements for mission concepts such as  AXIS and Lynx  are probably too lax to assure that spectral resolution requirements can be met. While the measurements reported here were made with a CCD, we note that they have implications for the performance of high aspect-ratio silicon active pixel sensors as well. 
\end{abstract}


{\noindent \footnotesize\textbf{*}Beverly J. LaMarr,  \linkable{lamarr@mit.edu} }
\footnote[0]{$^{\dagger}$ Deceased.}
\begin{spacing}{2}   

\section{Introduction}
\label{sect:intro}  
Large format, megapixel solid-state  image sensors have been mainstays of soft (0.1--10 keV) X-ray astronomy for decades. \cite{asca,acis,xmm,swift,suzaku,erosita} These devices  provide relatively large  fields of view, more than adequate spatial resolution, and moderate energy resolution.  Their imaging and spectroscopic capabilities also allow  discrimination between X-rays from cosmic sources and unwanted background generated by  charged particles encountered in the space environment.
\cite{asca,acis,xmm,swift,suzaku,erosita}

Future X-ray observatories are expected to require image sensors that exceed the capabilities of current flight detectors in a number of respects. For example, both the Lynx X-ray observatory \cite{Gaskin19}, a large mission concept studied by NASA, and the smaller  AXIS Probe-class concept\cite{AXIS}, would require imaging detectors with better spatial resolution (pixel size \mic{16}) than current systems, as well as excellent spectral resolution and quantum efficiency over a broader energy range  (0.2--12 keV)  (see Table \ref{tab:axlyreq}). Together these requirements dictate (for a silicon detector, at least) an unusually `tall' pixel, with pixel thickness to pixel width ratio $\sim$ 6:1, compared with current flight systems for which  that ratio is roughly $\sim$ 3:1.  Taller pixels entail greater lateral diffusion of the charge packets produced by absorption of X-ray photons, especially for X-rays at the low-energy extreme of the passband.  These soft photons are  absorbed very close to the detector entrance surface, so the charge packets they produce must drift farthest before collection,  and are therefore most likely to be shared amongst two or more adjacent pixels. Given  the relatively small total quantity of charge in these packets, good detection efficiency and accurate spectroscopy  at these energies requires low read noise.  Detector noise requirements for these missions must account for this phenomenon, and accurate modelling of detector performance requires knowledge of it. This challenge faces all silicon X-ray image sensors, including active pixel sensors of any architecture as well as CCDs.

\begin{table}[ht]
\begin{center}

\caption{Requirements for future X-ray Probe and Flagship missions} \label{tab:axlyreq}
\begin{tabular}{lcc}\hline \hline
Parameter & \multicolumn{2}{c}{Requirement}\\
&  AXIS\cite{AXIS} Probe & Lynx\cite{Gaskin19} Flagship \\ \hline
Angular resolution$^{1}$ & $0.5"$ & $0.5"$ \\
Energy range & $0.2$ -- $12$ keV & $0.2$ -- $10$ keV \\
Spectral resolution${^2}$ & $60$ eV FWHM @ 1 keV & 70 eV FWHM @ 0.3 keV \\
Pixel size & \mic{16} & \mic{16} \\
Read noise  & $\le 4$ electrons RMS & $\le 4$ electrons RMS\\
\hline
\multicolumn{3}{l}{\footnotesize{ Notes: 1. Half-power diameter on axis. 2. Full width at half maximum}}\\
\end{tabular}

\end{center}
\end{table}

Charge diffusion affects silicon sensor performance in other wavebands. In the visible and IR, its most important effect is a wavelength-dependent contribution  to the system point spread function~\cite{Groom,High2007}. In X-ray photon counting spectroscopic imaging applications, however, the effect of diffusion on spectral resolution and detection efficiency is of greater significance. We have been motivated to revisit this subject by the requirements of ambitious future X-ray observatories. The Lynx X-ray observatory \cite{Gaskin19}, and even smaller-scale missions such as the AXIS Probe-class concept\cite{AXIS}, require imaging detectors with moderate spatial resolution (\mic{16}), but this must be accompanied by good response over a very broad energy range (0.3--10 keV). Together these requirements dictate (for a silicon detector, at least) an unusually `tall' pixel, with pixel size to thickness ratio $\sim$ 1:6. Critically, Lynx studies of the high-redshift universe also require  excellent  detection capability at the lowest energies. Design of detectors meeting these requirements requires careful optimization of diffusion and readout noise. This challenge faces all silicon image sensors, including active pixel sensors of any architecture as well as CCDs. 

We note that other future developments in X-ray astrophysics may also require a better understanding of charge diffusion in solid-state image sensors. These include X-ray polarimetry \cite{H_marshall} and very high-resolution X-ray imaging. \cite{Schattenburg}.

We have been developing advanced CCDs for use in Lynx \cite{Bautz19}. CCDs are one of three types of imaging sensor, along with hybrid \cite{Falcone2019,Hull2019}  and monolithic\cite{Kenter2019} CMOS active pixel detectors, considered for this mission.   Our work, which also has potential applications in probe-class, Explorer and small missions, has focused on demonstrating fast, low-noise output amplifiers, low-power charge-transfer clocking, and development of appropriate application specific circuits \cite{Sven}. One of the devices we have tested features a relatively small pixel (\mic{8}) and a depletion thickness of approximately \mic{50}, and thus also offers the opportunity to evaluate the interplay of pixel size and charge diffusion in a sensor with a pixel thickness to width pixel-size  ratio near that required for Lynx and AXIS. In this paper we report measurements of charge packet size distributions as a function of device bias and X-ray energy obtained with this device. We compare these measurements to simulations made with a publicly available code \cite{Lageetal2021} developed to characterize the detectors of the Vera C. Rubin Observatory (VRO)\cite{VRO19}. We note that VRO CCDs are also high-aspect ratio devices, and that 5.9 keV X-ray characterization of these devices has been reported\cite{Kotov15}.  Our results have implications for the soft X-ray response of silicon active pixel sensors as well as CCDs.

\section{Data Acquisition and Analysis}

\subsection{Detector and X-ray Sources}
\label{sec:det}

Data reported here were obtained from a small-pixel, back-illuminated frame transfer detector developed at the MIT Lincoln Laboratory and designated CCID93. The  device features 512 x 512  pixels in both image and frame transfer areas. Pixels are   8-\mic{}  wide and the device is  \mic{50} thick.  Its gate structure  is implemented in a single layer of polysilicon with small inter-electrode gaps produced by photolithography. This configuration permits fast charge transfer with modest ($< 3$ V) clock swings \cite{Bautz19}. The device architecture allows the substrate to be biased independently of the gate voltage and channel stop potentials  to ensure that it is fully depleted. By varying the substrate bias voltage (hereafter `\vs') one can  change the electric field strength in the depletion region and thus change the amount of lateral diffusion of X-ray induced charge packets.

This test device was equipped with buried channel `trough' implants of varying widths, as well as a control region with no troughs. The trough implants provide significantly better charge transfer efficiency, and unless otherwise noted, we excluded data obtained in the trough-free region of the device. Details of the device architecture are described elsewhere\cite{Bautz19}.

We operated the detector at a temperature of $-50 ^{\circ}$ C with serial and parallel register rates of 2.0 MHz and 500 kHz, respectively and gate voltage swings of  $3.0$ V peak to peak. Readout noise is typically 4.5 electrons, RMS, under these conditions.  An STA Archon commercial controller  \cite{archon} was used to operate the detector and digitize the data. 

Data were collected in X-ray photon-counting mode using either a radioactive $^{55}$Fe source (producing Mn K X-rays at 5.9 and 6.4 keV) or a grating monochromator fed by an electron-impact source at energies below 2 keV.  The spectral resolving power of the monochromator is typically $\lambda / \Delta\lambda = \mathrm{E} / \Delta\mathrm{E} \sim$\ 60--80, far exceeding that of the detector at the energies of interest. In this work, we focus on results at two energies, 5.9 keV and 1.25 keV from Mn K$\alpha$ and Mg K$\alpha$ photons, respectively.

\subsection{Data Acquisition}

Data were obtained for each energy and substrate bias value of interest by repeatedly  reading the detector, storing each full-frame readout as a distinct file. The exposure time for each frame was 0.15 s, and source flux was adjusted to  less than $\sim 25$ detected photons (events) per frame ($\lesssim 10^{-4}$ events per pixel)  to minimize pileup. Approximately $2.3\times 10^{5}$ events were collected for each configuration. Data were obtained at seven values of \vs \   ranging from $-$0.2 V to $-$20 V.

\subsection{Analysis Methods}
\label{sec:anl}

Data were acquired in groups of 100 consecutive frames. From each acquired frame  an average of the overclock region was subtracted  to remove drift of the DC level.  A bias (zero-signal) frame was then computed for each group using a clipped average algorithm to remove signal from  X-rays and cosmic rays. The bias frame  was subtracted pixel-by-pixel from each data frame in its group. X-ray events were then identified by searching the bias-subtracted frames for pixels with amplitudes exceeding a fixed threshold (the `event threshold') which are also local maxima. The event threshold was set to 10 times the RMS noise level for 5.9 keV X-rays, and 7.5 times the noise for 1.25 keV.  This level is high enough to reject spurious events due to noise while low enough to accept evenly split x-rays. For each  event the values of a 7$\times$7 pixel array centered on the located maximum were recorded in an event list. 

We found that in this relatively small pixel device almost all of the events have signal charge spread over multiple pixels.  To characterize the signal distribution we fit\cite{mpfit}  a two-dimensional  Gaussian function to the pixel values of each event. In these fits the Gaussian standard deviation ($\sigma_{spatial}$) was constrained to be the same in both dimensions, and a constant additive term was included to allow for a local offset bias level. In this way we obtained five parameters (two for  position, plus one each for amplitude, $\sigma$, and bias offset) for each event.

For comparison, we also computed the sum of all the pixel values in a seven by seven pixel island above a second, lower, threshold value, known as the `split event' threshold. This is the event energy reconstruction method in use for all past and current X-ray CCD missions for astrophysics, but with the number of pixels in each island increased to compensate for the much smaller pixels in our device.  We set the split threshold at four times the RMS readout noise per pixel.  


\subsection{Simulations}
\label{sec:simulation}

Simulated event lists were constructed using \textsc{Poisson CCD} \cite{Lageetal2021}, a software package that, given the structural and operational characteristics of the detector, (1) solves Poisson's equation for the electric field in three dimensions at all points within the detector volume, and (2) tracks the drift and diffusion of electrons placed within this volume until they reach the buried channel, where they are collected in pixels defined by the channel stop and barrier gate fields. We set up the simulated detector to reflect as nearly as possible the characteristics and operating conditions of the CCID93 device described in Section \ref{sec:det}: 50-\mic{} thickness, 8-\mic{} pixels, and 3-phase gate structure with one collecting gate held at +1.5 V and two barrier gates held at $-$1.5 V. Implant and doping parameters were provided by MIT Lincoln Laboratory. The simulation volume covered 9$\times$9 pixels, with non-linear grid spacing in the vertical direction allowing finer grid sampling of 15 nm at the front and back sides of the device to properly capture the electric field structure, and coarser grid sampling (up to 150 nm) throughout the bulk of the device. The electric field was simulated at an operating temperature of $-50 ^{\circ}$ C at each of the \vs\ settings used for the real device.

For each \vs, we simulated detection of 200,000 5.9 keV and 1.25 keV photons by introducing small clouds of electrons and allowing them to drift and diffuse until collected in pixels. The number of electrons in each cloud was drawn from a Fano noise distribution appropriate for the photon energy in Si (e.g., 1615$\pm$14 electrons for 5.9 keV), and the interaction depth was drawn from an exponential distribution with appropriate attenuation length. Each interaction location was generated from a uniform distribution across the central pixel of the 9$\times$9-pixel simulation volume. The final pixel distribution of electrons was converted into an event island in energy units. Simulated Gaussian readout noise with RMS of 4.5 electrons was added to each pixel, and event detection and characterization were performed in the same way as the lab data.

The default charge diffusion parameters in \textsc{Poisson CCD} produced much more diffusion in the simulations than we observe in the lab data, with much larger event sizes. Since the purpose of the simulations was to illuminate the physical processes responsible for the observed data, we tuned the \textsc{DiffMultiplier} parameter so that the simulated event size distributions closely matched the real data, as described in the following Section. These \textsc{DiffMultiplier} values ranged from 1.4 to 1.7 for \vs $= -$0.2 to $-$20 V, compared to the value of 2.3 determined from charge diffusion measurements in Vera Rubin Observatory CCDs\cite{Lageetal2021}. This parameter is implemented in \textsc{Poisson CCD} as a scale factor for the charge carrier thermal velocity, given by $v_{th} = (8kT/m_e\pi)^{1/2} \times \mathrm{DiffMultiplier}$,\cite{Lageetal2021} where $m_e$ is the bare electron mass. In effect it specifies a value for the thermal velocity effective mass of the electron in the Si conduction band, $m^*_{e,tc}$. Previous estimates suggest $m^*_{e,tc}/m{e} \approx$ 0.27--0.28\cite{Green1990}, which would indicate \textsc{DiffMultiplier} $\approx 1.9$ as an appropriate value. The modest difference between that and the value \textsc{DiffMultiplier} $=1.7$ required to match our data at highly negative \vs\ (fully depleted substrate) is likely due to the limitations of the carrier transport model used in \textsc{Poisson CCD}\cite{Lageetal2021}. The larger differences at less negative \vs\ could arise from incorrect treatment of undepleted bulk in the simulator (C.~Lage, private communication). We saw no difference in the amount of diffusion if electron cloud Coulomb repulsion was turned on (\textsc{Fe55} mode). These issues will be explored in a future paper aimed at further validating the simulations and producing a higher-fidelity simulation methodology for these thick, small-pixel X-ray CCDs. We will also consider implementing techniques used to model drift and diffusion developed recently by other groups\cite{Haroetal2020,Prigozhinetal2021}.

\section{Measurements}
\subsection{Overview}
\label{sec:overview}
\begin{figure}
\includegraphics[width=.95\linewidth]{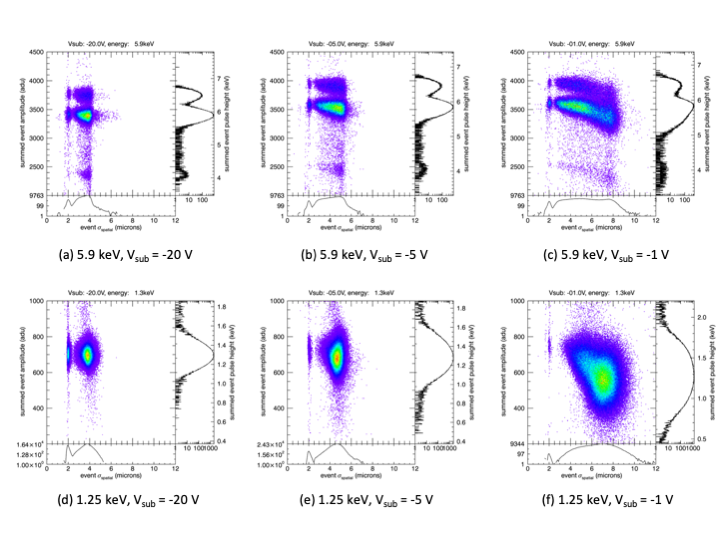}

\vspace*{\baselineskip}

\caption{Scatter plots of summed-pixel amplitude vs.\ width (2D Gaussian $\sigma$) for X-ray events at two energies and three values of the substrate bias (\vs). The color indicates the areal density of events in the width/amplitude plane. Histograms of event amplitude and width are projected onto the right and bottom edges, respectively, of each panel.}
\label{fig:fe55_amp_sig}
\end{figure}

We measure (two-dimensional) position, amplitude, and width (2D Gaussian $\sigma_{spatial}$) of each event as described in Section \ref{sec:anl} above. Figure \ref{fig:fe55_amp_sig} shows scatter plots of summed-pixel event amplitude vs. event width for both X-ray energies and three different values of internal detector bias \vs.  Here color indicates density of events in the amplitude-width plane, and projections of the data in these two axes show the corresponding distributions in amplitude and width. Approximately 230,000 events are represented in each panel of the figure.  

The (vertical) amplitude  distributions show the detector's spectral response function at energies of 5.9 and 6.4 keV (upper panels) and 1.25 keV (lower panels). Silicon K-escape and fluorescence lines are evident in the upper panels. The (horizontal)  width distributions reflect the amount of lateral diffusion experienced by charge packets before collection. 

Comparison of the three columns  in Figure \ref{fig:fe55_amp_sig} shows how the device response changes with the strength of the electric fields in the detector's photosensitive volume, with the strongest field  (\vs$ = -20$V) on the left and the weakest (\vs$ = -1$ V) on  the right.  Thus in the left column  the relatively strong fields provide relatively rapid charge collection, and thus relatively small event widths, as well as relatively good spectral (amplitude) resolution. As  field strength decreases in the center and right columns, the figures show the effects of  progressively smaller fields ( \vs $ = -5$V and $ -1$V, respectively): both width and amplitude distributions become progressively wider. In fact, the double-peaked width distribution in Figure~\ref{fig:fe55_amp_sig}c suggests that the detector is no longer fully depleted at this low value of substrate bias. Events with the largest widths at this value of \vs \ also show depressed amplitude, implying that events produced in the undepleted region of the detector suffer from incomplete charge collection as well as large lateral diffusion~\cite{GYP2003, Miyata02}.

\begin{figure}
\includegraphics[width=0.95\linewidth]{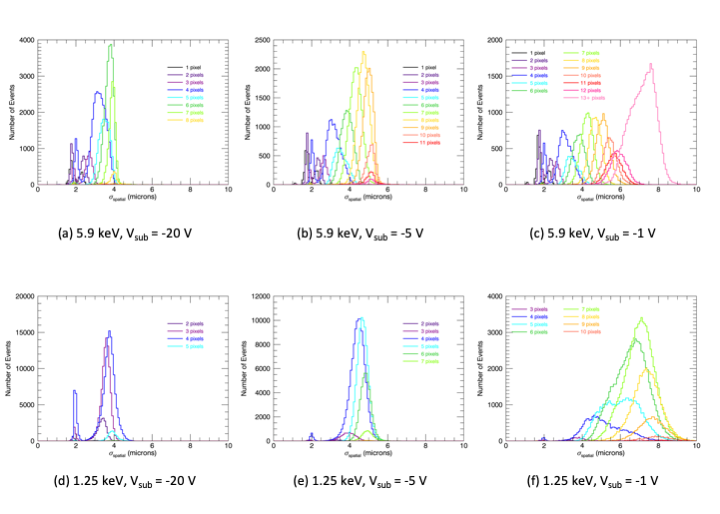}

\vspace*{\baselineskip}

\caption{Histograms of fitted  event width (Gaussian $\sigma_{spatial}$) by number of pixels above the split threshold (pixel multiplicity) for various values of the substrate bias \vs \  and for two X-ray energies.}
\label{fig:fe55_sig_num}
\end{figure}

A closer look at the spatial width distributions is provided in Figure~\ref{fig:fe55_sig_num},  which shows  distributions for events comprising various number of pixels above the split threshold. (Hereafter we denote the latter quantity as `pixel multiplicity'). The panels in this figure correspond to those in Figure~\ref{fig:fe55_amp_sig}.

Comparison of the upper and lower panels of Figure~\ref{fig:fe55_sig_num} shows the marked energy dependence in the size distributions. Generally, events with a given pixel multiplicity are spatially smaller at the higher energy. In addition, for 5.9 keV events there is a clear correlation between width and number of pixels above threshold, while at 1.25 keV the variation in size with number of pixels is much smaller. Finally, at the stronger fields (\vs$ \leq -5$V), the maximum size of events is similar at both energies (about \mic{4} for \vs $=-20$V). 

These observations are straightforward consequences of the expected energy dependence of X-ray  interaction depths in the detector, coupled with the  dependence of lateral diffusion on interaction depth. The relatively more penetrating 5.9 keV X-rays (attenuation length in silicon approximately \mic{29} ) are absorbed throughout the \mic{50} thickness of the detector.  The 1.25 keV X-rays (attenuation length \mic{5}) all interact in the third of the detector closest to the entrance window. As a result, there is a much larger range of lateral diffusion and therefore sizes in the population of  5.9 keV events. An important conclusion from this figure, however, is that when the detector is fully depleted (two left columns), the maximum amount of diffusion is the same at both  energies. 

In all cases the histogram peaks near  $\sigma =$ \mic{2} are associated with events for which charge is detected in four or fewer pixels. Although the amplitude of these events can be measured accurately by summing the pixels above split threshold, their sizes cannot be resolved by our detector or reliably measured using the Gaussian fitting algorithm.
Since our pixel size is comparable to or greater than than the intrinsic (`pre-pixelization') size of the charge packets, our best-fit values of $\sigma_{spatial}$ are biased high relative to those of the intrinsic distributions.  The magnitude of this bias depends on the intrinsic width.  Analytic calculations suggest, and simulations confirm,  that this bias ranges from more than 30\% to less than 10\% for \mic{3} $\lesssim  \sigma \lesssim $ \mic{6}. For $\sigma < $ \mic{3}, our fitting algorithm  fails to converge, and tends to return values of $\sigma \approx $ \mic{2}.  This accounts for the small peaks at this value in the width distributions shown  in Figures~\ref{fig:fe55_amp_sig} and ~\ref{fig:fe55_sig_num}.

Such spatially unresolved charge distributions can arise in different ways.  At 5.9 keV,  some X-rays  can penetrate deep into the detector to interact close to the CCD's buried channel, and thus can produce events that suffer very little diffusion. After pixelization all information about the intrinsic size of these  events is lost and the Gaussian model does not describe them well. At 1.25 keV, all events must be subject to considerable diffusion. In this case,  however, the relatively large spatial extent of the charge distribution, coupled with  the relatively small total number of electrons, can produce low signal-to-noise ratios in the peripheral pixels. Simulations confirm that fits may not converge in these circumstances.  


%

\FloatBarrier

\subsection{Measurement and applications of the event charge distribution}
\label{sec:shape}
The data presented in the previous section illustrate the broad range of event sizes that occur even under monochromatic X-ray illumination. A complete characterization of detector response at even a single energy thus requires knowledge of both  the distribution of charge for events of a given width (which we have assumed to be Gaussian in the foregoing), and the distribution of event widths. Previous work by Prigozhin and co-workers\cite{GYP2003} has shown how this problem can be addressed by direct measurement. 

\begin{figure}[t]
\begin{center}
\includegraphics[width=0.8\linewidth]{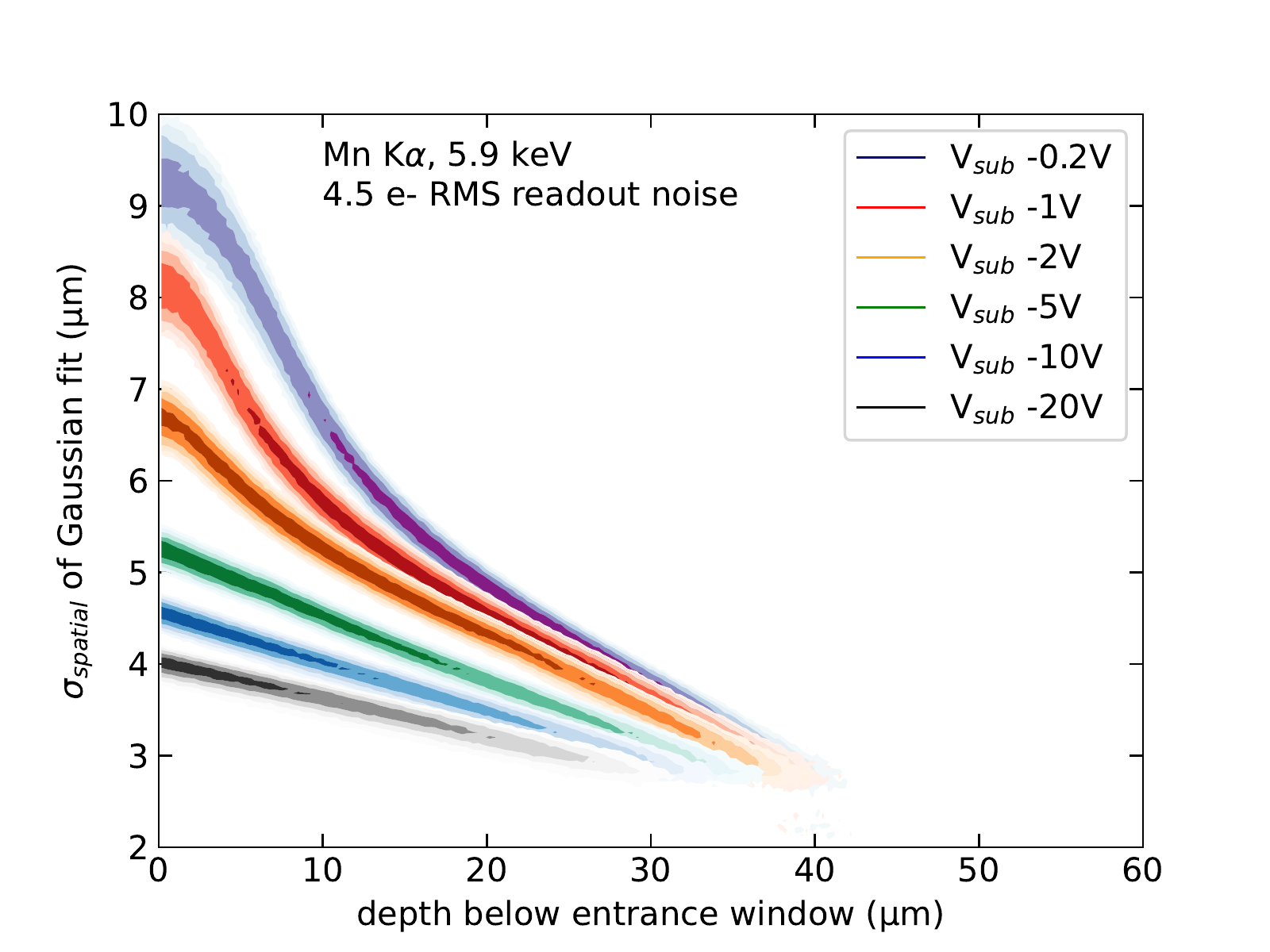}
\end{center}
\caption{Simulated best-fit Gaussian $\sigma_{spatial}$ as a function of the photon interaction depth for different values of substrate bias parameter \vs\ and photon energy of 5.9 keV. The device is \mic{50} thick. Contours show 5\%, 10\%, 20\%, 40\%, and 70\% of the maximum density of points for each \vs. There is a clearly monotonic trend of event width with depth. It is also clear that  the maximum event size depends on \vs.}
\label{fig:SimSigmaDepth}
\end{figure}

Briefly,  we can exploit the fact that for a monochromatic incident beam, both the number of X-rays absorbed and the magnitude of lateral diffusion are monotonic functions of the depth in the detector  at which the X-ray is absorbed. The former relation is simply a consequence of the attenuation of the X-ray intensity with depth, and can be determined reliably from known material properties. The latter results because the time required for the liberated electrons to drift from the interaction point to the buried channel, and thus the amount of lateral diffusion, decreases with depth. The expected width-depth relation is illustrated in Figure~\ref{fig:SimSigmaDepth}, which shows simulation results for 5.9 keV X-rays  for various values of \vs. Because (for a given field configuration) this relationship is also monotonic in depth, it is possible to measure it using X-ray data. Moreover, it is also possible to obtain a high-quality measurement of the shape of the charge distribution for events interacting at a given depth, and thus test our assumption that this distribution is Gaussian.


\FloatBarrier

\subsubsection{Average charge distribution at fixed depth}
To characterize event charge distributions at fixed depth we  first order events by their measured values of  $\sigma_{spatial}$, and then select 1000 events surrounding each of several percentile points in the cumulative $\sigma_{spatial}$ distribution, thus selecting a group of events coming from approximately the same depth in the device. To minimize the effects of pixelization, we create a 70$\times$70 subpixel grid over the 7$\times$7 pixel island around each event's central pixel. Each element of the subpixel grid as assigned an amplitude equal to 1/100 of the amplitude of the pixel in which that element lies.  We next align the centroid of each event with the origin of the subpixel grid, and then average all subpixel amplitudes over all events in the percentile group. Since  centroids can be determined with precision finer than a single pixel, this procedure produces a mean event charge distribution with  subpixel resolution. 

\begin{figure}[h]
\includegraphics[width=0.95\linewidth]{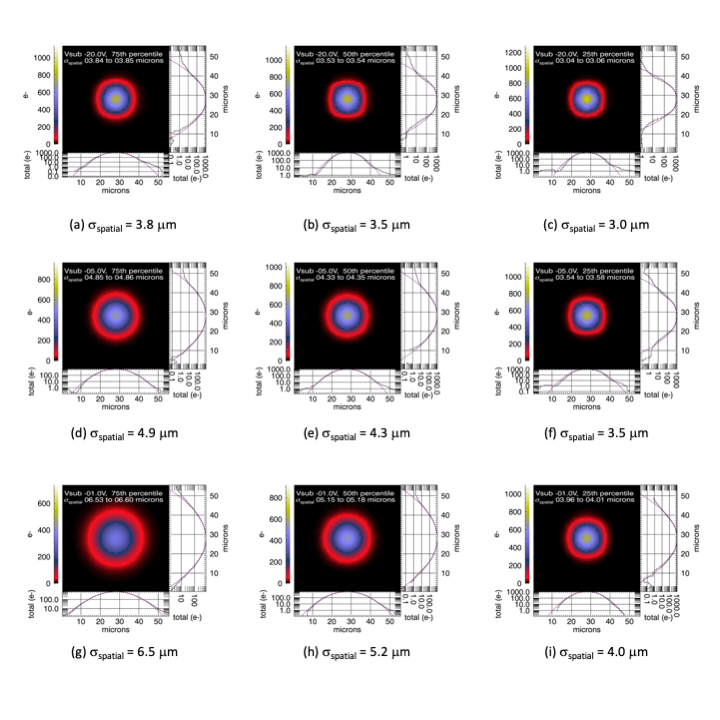}

\vspace*{\baselineskip}

\caption{Measured average event profiles produced by 5.9 keV X-rays.  The top, middle and bottom  rows show results for \vs $ = -20$ V, $-5$ V and $-1$ V, respectively.  The left, middle and right  columns  are for events in the 75th, 50th and 25th width percentiles, respectively. Horizontal and vertical Projections of the two-dimensional distributions, together with best-fit Gaussians, are shown for each case. Here the units of ordinates are electrons per pixel. 
}
\label{fig:avgEv}
\end{figure}

The results for  5.9 keV at three different percentiles  and three different values of \vs\ are shown in Figure~\ref{fig:avgEv}. Each row corresponds to a different value of \vs, with the strongest field at the top.  Each column is a different value of percentile in the cumulative width distribution, with the largest  percentile width at the left.

Qualitatively, the figure shows as expected that for a given field distribution, event widths become smaller as the interaction depth increases (to the right), and that for a given depth, event widths also become smaller as the field strength increases (bottom to top). Note that at \vs $=-1$ V (bottom row) the detector is only partially depleted,  and the cloud size is quite large. In both the middle and top rows,  with more more negative \vs \ the device is fully depleted, and cloud size shrinks accordingly. As noted in Section~\ref{sec:overview}, at very small cloud sizes ($\sigma_{spatial} < $ \mic{4}), the effects of pixelization become evident:  the inferred shape  tends to become non-circular, reflecting that of the pixels rather than that of the intrinsic charge distribution.   We shall return to the consequences of this  under-sampling in Section~\ref{sec:amp} below.

How accurate is our assumption of a Gaussian charge distribution?  Projections (i.e., sums along rows and columns) of the signal distribution are shown to the right and beneath each panel in Figure~\ref{fig:avgEv}, along with best-fit Gaussian curves. The central portions of the profiles are clearly very well described by a Gaussian function over at least two orders of magnitude, which is consistent with theoretical calculations for clouds formed within the depleted region\cite{PavlovNousek99}.  There is a slight excess in measured signal compared to the Gaussian wings, but the level of discrepancy is on the order of 0.1\% of the total charge packet signal per pixel. We note that the deviation is largest for the data with the smallest $\sigma_{spatial}$, which are  most affected by the pixelization.  We  also note that for vertical signal distributions the tail on the top side is noticeably higher than on the bottom side, no doubt as a result of charge transfer inefficiency  that causes electrons trapped during parallel transfer to be re-emitted into the pixels behind the event center.  A similar distortion of charge-cloud shape has been used previously as a diagnostic of charge transfer inefficiency in electron multiplier CCDs~\cite{Kotov2021}. We conclude that charge packets are Gaussian to a very good approximation, and that the most significant deviations arise at small widths as a result of pixelization. 
\FloatBarrier

\subsubsection{Inferring interaction depth from the cumulative event width distribution}
\label{sec:depth}
As noted at the beginning of this section,  since both the density of X-ray events and the size of the corresponding charge distributions  vary monotonically with interaction depth, it is possible to map the density of the event widths onto the distance from the illuminated surface. This idea was proposed previously by Prigozhin and collaborators\cite{GYP2003}. We implement this map as follows. The integral number of  photons absorbed between the illuminated detector surface and the plane at depth $z$ below that surface) can be written
$N_z = N_0 (1-exp(-z/{\lambda}))$. Here $N_0$ is the incident photon fluence. The total number of events $N_{total}$ absorbed in  an ideal device with thickness $t$ is $N_{total} = N_0 (1 - exp(-t/\lambda))$, so the cumulative fraction of events absorbed above depth $z$,  $N_{z\_frac} = N_z/N_{total}$,  is  
\begin{equation}
N_{z\_frac} = \frac{1 - exp(-z/\lambda)}{1 - exp(-t/\lambda)}
\label{eq:cum_depth}
\end{equation}

\begin{figure}[h]
\includegraphics[width=1.0\linewidth]{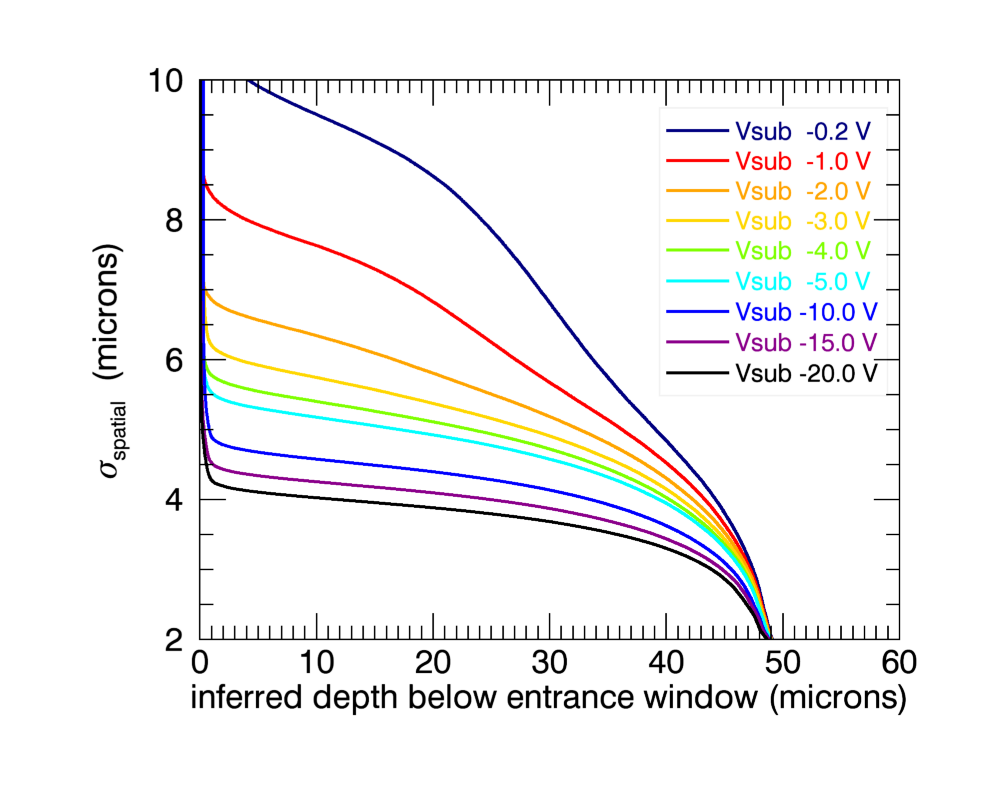}
\caption{
Measured cumulative distributions of event width ($\sigma_{spatial}$) for 5.9 keV X-rays as a function of absorption distance (depth) below the entrance window.}
\label{fig:SigmaDepth}
\end{figure}

This relationship is monotonic and can therefore be inverted to give  the value of $z$ as function of the fraction of events above that depth:
\begin{equation}
 z = -\lambda \ ln(1 - N_{z\_frac}(1 - exp(-t/z))) 
 \label{eq:depth}
\end{equation}
Both the detector thickness ($t=$ \mic{50}) and, for a given  photon energy, the absorption length ( $\lambda=$ \mic{29} for  5.9 keV X-rays) are known so the relationship between $z$ and $N_{z\_frac}$ is completely specified. 

We  also sort the observed events in order of width to form their cumulative  distribution as a function of that parameter. Invoking our physically motivated assumption that width decreases with depth, we    use this empirical width distribution, together with equation~\ref{eq:depth},  to determine the relationship between width and depth.  We note that it is important to include all events in this calculation, even the ones with very small measured width, in spite of the fact that the width of such events is not measured accurately, as discussed in Section \ref{sec:overview}. 

The resulting measurements  of event width as a function of  depth below the entrance window are shown for different values of \vs \  in Figure~\ref{fig:SigmaDepth}.
As expected, stronger internal fields (more negative \vs) reduce the  diffusion time and thus event widths. At \vs $>-3$  V the shape of the curves is different than at lower \vs. This is almost certainly due to formation of an undepleted layer of silicon near the illuminated surface, in which charge collection may be inefficient, excluding some events from our analysis. This would violate our assumption that all events are included, and would introduce errors in our assignment of depth to width. This interpretation is supported by the simulated width-depth curves shown in Figure~\ref{fig:SimSigmaDepth}; the simulations do not include such surface effects, and the simulated width-depth curves do not show this shape change. 

Another limitation of this  algorithm is that it neglects uncertainties in   measurements of $\sigma_{spatial}$.  We defer investigation of the significance of this effect to future work. 

\subsubsection{Detector characterization from differential event width distributions}
\label{sec:widthvbias}

In this section we turn from the cumulative to the differential form of the  width distribution and  show how it varies with detector bias. In principle, differential width distributions are an important tool in predicting detector performance. They can be used, for example to estimate proportions of events with different pixel multiplicities, which in turn are needed  to  predict the  spectral resolution of an image sensor. They also provide powerful observables for use in validating a detector model, especially when measured over a range of internal field conditions.  

\begin{figure}[ht]
\begin{center}
\begin{tabular}{c}
\includegraphics[width=0.8\linewidth]{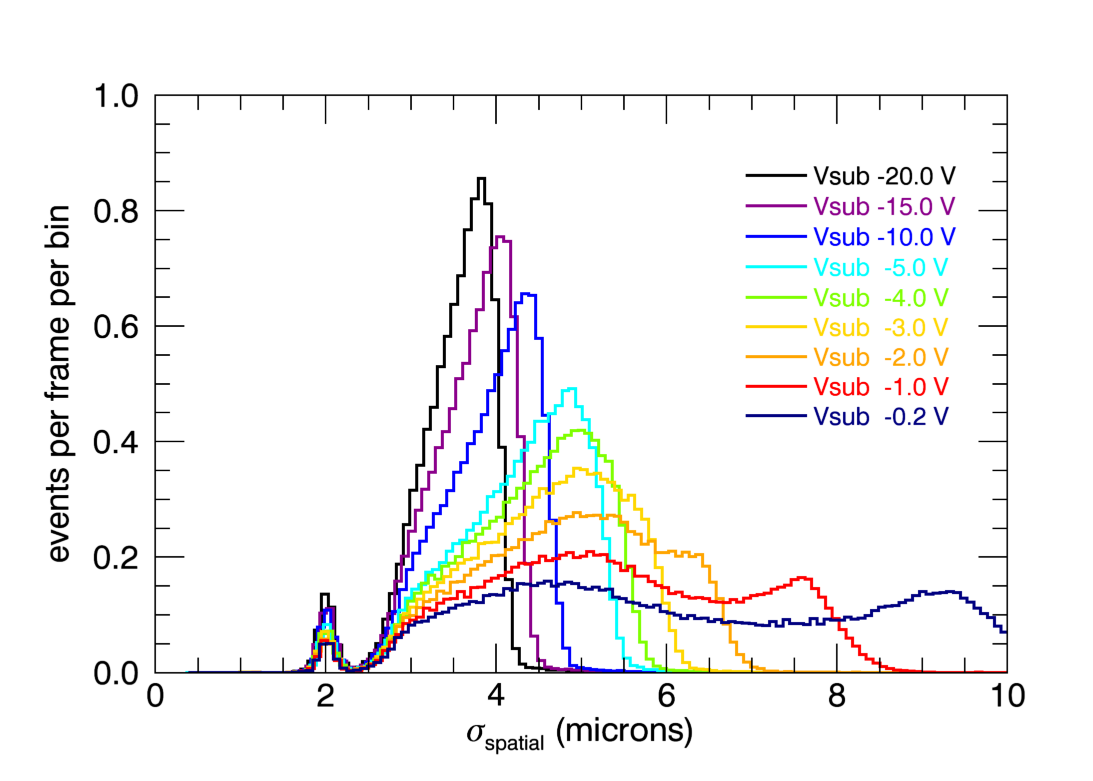}
\end{tabular}
\end{center}
\caption 
{ \label{fig:SigmaGE4}
The differential width distributions  of  events with pixel multiplicity greater than 3 produced by 5.9 keV X-rays. Distributions for a range of detector bias (\vs) conditions are shown.  } 
\end{figure}

 Figure~\ref{fig:SigmaGE4} shows these distributions for 5.9 keV X-rays over a range of substrate bias. The distributions shown include only those events with measurable widths, that is, those for which  at least four pixels have measurable charge. 

The bias dependence of these distributions tells a now familiar story, but also provides new insights. At the largest bias (\vs $= -20$ V, black histogram) the bulk of the silicon is fully depleted. As there is a strong electric field throughout the device, drift times are short,  so  the corresponding width distribution  is relatively narrow, and shows  a sharp edge at \mic{3.9}. This edge can be interpreted as the width of events interacting in the immediate vicinity of the detector entrance surface. This maximum event width should apply to events of all energies, and is thus a very useful parameter for estimating the response  of the detector to low energy photons which are necessarily absorbed there. The shape of the left side of the distribution  is determined by the width-depth relation, and is thus sensitive to internal field distributions and charge transport properties of the detector. The steep but finite slope of the edge is in principle a measure of the statistical uncertainty of the width measurement. The small peak near $\sigma_{spatial} = $ \mic{2} is caused by very narrow events  produced close to the buried channel, for which our width measurement is not reliable, as discussed above in Section \ref{sec:overview}.

As \vs \ increases (algebraically), the internal  electric field strength drops, broadening the distribution to larger event widths, until  a clear transition occurs at \vs $=-4$V. This is caused by the formation of an undepleted region near the illuminated surface of the device. As internal field strength drops further, the events in the undepleted region form a separate hump in the histogram at much larger widths.  In this way the differential width distributions provide a readily measurable indicator of full depletion within the device. A simple 1D calculation of \vs\ at which full depletion of \mic{50} thick slab of silicon with doping concentration of 2.65x$10^{12}$ cm$^{-3}$ would occur yields a value of \vs\ = 5.1 V, in good agreement with   both our experimental and simulated full depletion transition.

\begin{figure}[ht]
\begin{center}
\includegraphics[width=0.8\linewidth]{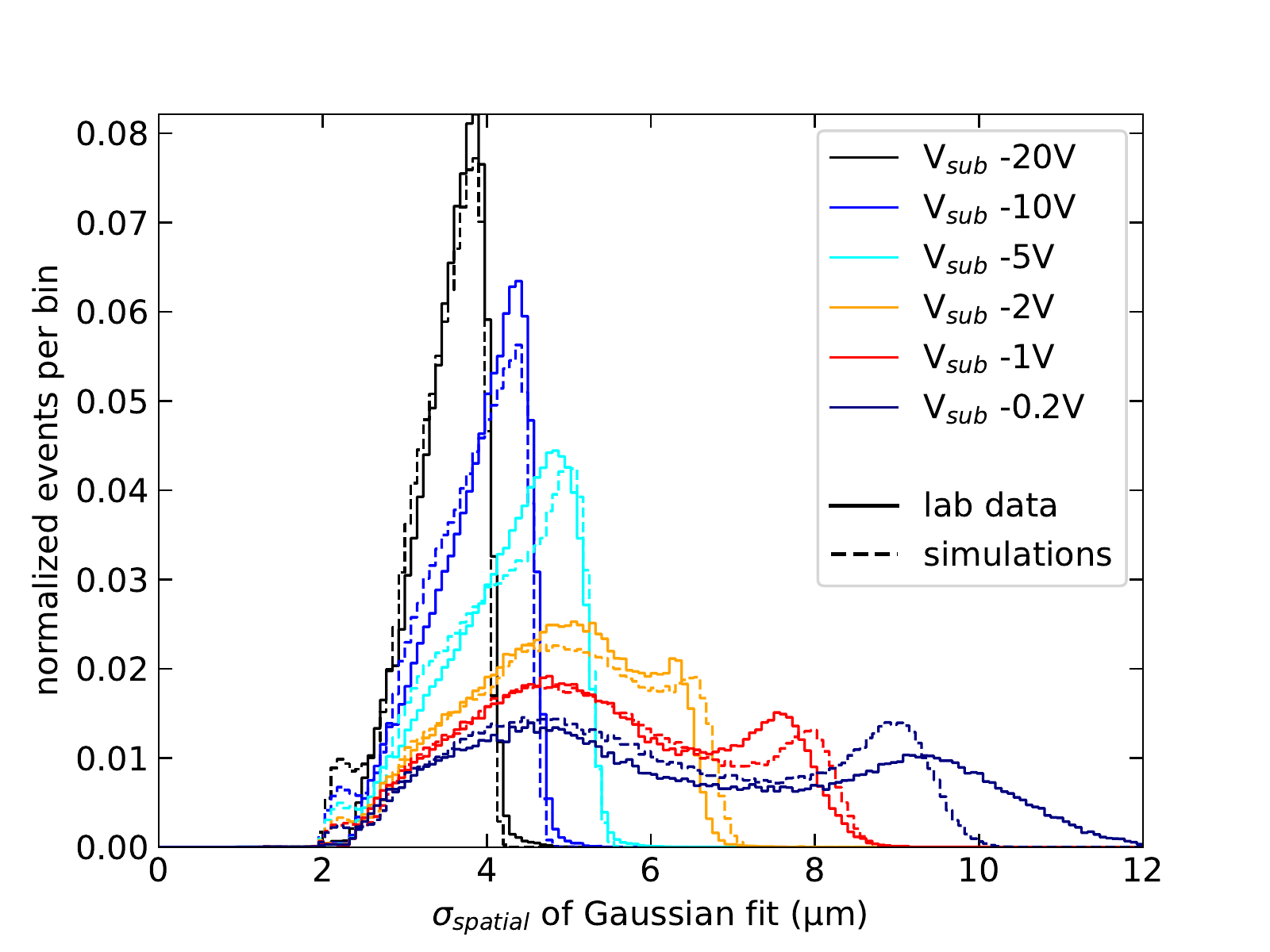}
\end{center}
\caption 
{\label{fig:SimSigmaGE4}
The distribution of Gaussian $\sigma_{spatial}$ from simulations of 5.9 keV X-rays, compared to the data (see Figure \ref{fig:SigmaGE4}). Only events with at least 4-pixel multiplicity are shown, and for clarity some \vs\ values are not plotted. A single diffusion factor was tuned for each simulation to recover a similar distribution to the data, as described in the text. The simulated distributions are similar to the data for large negative \vs\ in which the device is fully depleted. For small \vs\ and large $\sigma_{spatial}$, the distributions differ significantly, likely due to limitations of the simulations.} 
\end{figure}

Figure~\ref{fig:SimSigmaGE4} compares the event width distributions derived from the simulations with those from the measurements. We tuned the simulation's  \textsc{DiffMultiplier} parameter (described in Section  \ref{sec:simulation} above) for each \vs\ so that the peak of the simulated distribution matched the peak of the measured distribution. The tuned values have a small range of 1.4--1.7, with smaller values for smaller (less negative) \vs.  The upper end of this range is not too different from the value  \textsc{DiffMultiplier} $\approx 1.9$ expected for a cannonical value of the thermal velocity electron effective mass \cite{Green1990}, although it differs noticeably   from the value of 2.3 determined for (fully depleted) Vera Rubin Observatory CCDs\cite{Lageetal2021}. Our simulations for  \vs\ $< -5$ V, at which the device is fully depleted,  show good agreement with our data. Agreement is poorer at less negative \vs; the measured distribution for -0.2 V has a more extended tail to larger $\sigma_{spatial}$ than the simulations would predict. This is likely due to both a poor representation of the backside passivation layer and an inadequate correction for backside surface charge losses in the simulation. These issues are exacerbated when the region near the backside is not fully depleted, producing a field-free region where electrons can linger and greatly affecting the measured charge diffusion. Both the need for tuned diffusion and possible improvements to undepleted backside characterization will be addressed in a future paper focused on simulations.

\FloatBarrier

\subsection{Event Amplitude Estimation}
\label{sec:amp}

The spectral resolution of an X-ray image sensor is its spectral resolution, that depends  on the accuracy with which the total charge associated with  an  X-ray event (the event 'amplitude') is measured. Traditionally, in large-pixel devices (those with pixels much larger than characteristic event widths) used for Chandra, XMM-Newton, Suzaku and other X-ray instruments, the amplitude is estimated as the sum signal in a few pixels around a local maximum exceeding the split threshold.  In detectors with pixel sizes comparable to characteristic event widths,  the majority of events have charge spread over multiple pixels. In this case, numerous pixels that fall below the split threshold may in aggregate contain a significant fraction of the total charge, and it is natural  to consider whether alternative methods that explicitly account for finite event width could provide a better estimate  of event amplitude. 

\begin{figure}[p]
\includegraphics[width=.95\linewidth]{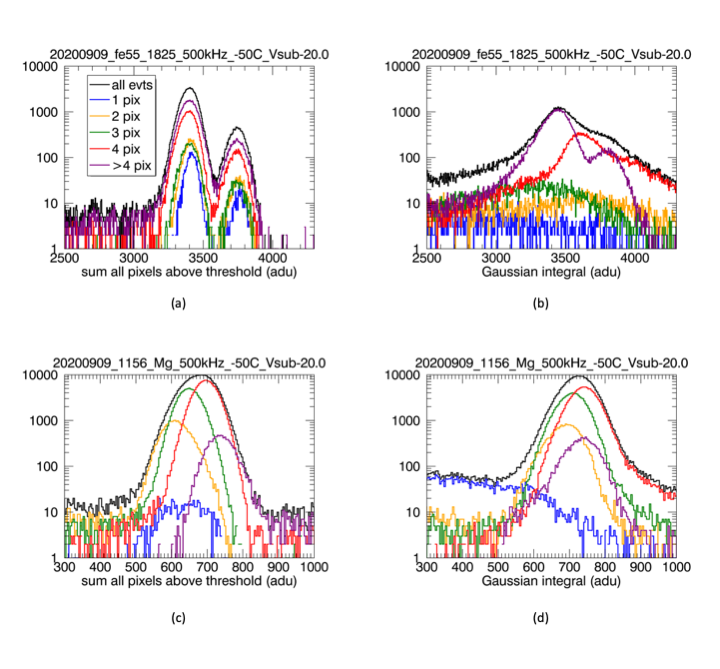}

\caption{Measured event amplitude spectra. Different colors show results for events of various pixel multiplicities.  Left and right panels show results for 5.9 keV and 1.25 keV, respectively.  Event amplitudes are calculated by summing all pixels above split threshold (top panels) or by integrating under the Gaussian fit to the spatial charge distribution of each event (bottom panels). All data were obtained with strong internal electric fields (\vs $ = -20$V). At the higher energy, the sum of pixels method produces a much narrower spectral response, although the performance of the Gaussian method can be improved by eliminating  events with low multiplicity. At the lower energy, the Gaussian method produces a similar core spectral response compared to the summed pixel method, but with an extended high-energy tail populated predominantly by  4-pixel events. The broad spectra of low-multiplicity events at both energies are caused by poor performance of the Gaussian fit for undersampled distributions.}
\label{fig:spectra}
\end{figure}

\begin{figure}[p]
\includegraphics[width=0.95\linewidth]{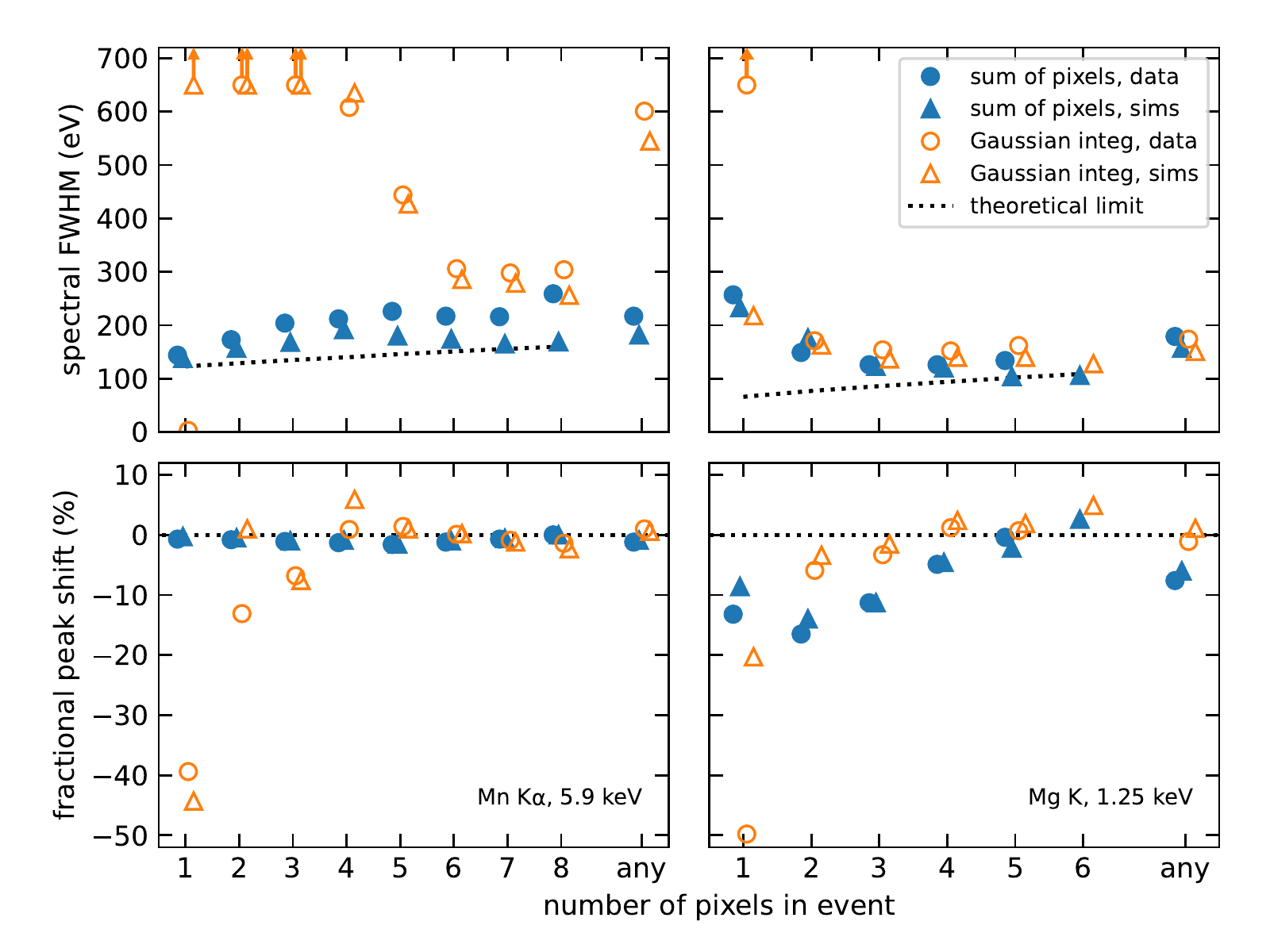}
\caption{Spectral response parameters FWHM (top) and peak shift (bottom) for 5.9 keV (left) and 1.25 keV (right). These were measured by fitting a single Gaussian to the spectra shown in Figures \ref{fig:spectra} and \ref{fig:simspectra}. Points with arrows indicate values out of the plot range. The spectral FWHM ``theoretical limit'' is the Fano spectral width convolved with pixel-based noise of 4.5 electrons RMS. At 5.9 keV, the spatial Gaussian integral summation method performs poorly compared to a simple sum of pixel values, except when the number of pixels above threshold is large. At 1.25 keV, the spatial Gaussian performs similarly or better than the pixel summation method. See Tables \ref{tab:specwidths} and \ref{tab:specpeaks} in Appendix \ref{sec:appendix} for a full tabulation of values.}
\label{fig:specresponse}
\end{figure}

We evaluated direct fitting of the event charge distribution  as a method for determining event amplitude.    A comparison of this approach with simple summing of pixels exceeding the split threshold is shown in Figure \ref{fig:spectra} for 5.9 keV and 1.25 keV photons.  To investigate the magnitude of charge lost due to the thresholding used in the traditional algorithm, we separately plot spectra for events of various pixel multiplicities.  In Figure \ref{fig:specresponse}, we show spectral FWHM and peak location as a function of pixel multiplicity (see also Appendix \ref{sec:appendix}). These data were obtained with substrate bias \vs $= -20$V, providing the maximum internal electric field strength. In general,  integrating  the best-fit spatial Gaussian does produce a slightly higher estimate for the event amplitude, consistent with the idea that a functional form accurately describing the event shape can recover signal lost to the surrounding pixels that fall below split threshold. On the other hand, the spectral distributions  derived from the Gaussian fits are noticeably broader and themselves clearly non-Gaussian,   especially at 5.9 keV.   

The 5.9 keV spectra in the left panels of Figure~\ref{fig:spectra} and the corresponding response parameters in the left panels of Figure \ref{fig:specresponse} show that the Gaussian fit performs worst for events for which a relatively small number of pixels (roughly 5 or fewer) exceed the split threshold. We attribute this behavior to the spatial under-sampling of events with intrinsically narrow charge distributions. We also note that at this energy there is only a small change in the spectral peak location with pixel multiplicity for either amplitude determination method, suggesting that relatively little charge is contained in pixels below the split threshold. 

The situation is different at the lower energy, as the right panels of Figures \ref{fig:spectra} and \ref{fig:specresponse} show.  Here the performance of the two amplitude determination methods is quite similar, and spectral widths are in some cases marginally better for  Gaussian fits. Remarkably, good results are obtained with this method even for events with as few as two pixels above threshold, suggesting that as expected these events are more extended and suffer less from under-sampling than their counterparts at higher energy.  The low pixel multiplicity of  these events is due to truncation by the threshold rather than an  intrinsically narrow spatial distribution. This interpretation also explains the systematic increase of spectral peak location with pixel multiplicity at this energy. Figure \ref{fig:specresponse} shows that Gaussian fits are indeed less susceptible, though not immune, to spectral broadening due to charge loss  at this energy: peak locations change less with pixel multiplicity, and, as a result, the spectral full-width at half maximum for the entire data set is actually slightly better for the fitting method than for pixel summation.  

A complication of this simple picture is presented by the spectrum of four-pixel events derived from Gaussian fitting, which shows a high-energy tail. We interpret this tail as another consequence of poor fitting arising from (maximal) under-sampling of those charge clouds with spatial centroids close to the  center of a 2x2 pixel array. Examination of the fitted centroids of these tail events confirms this interpretation.  We defer a detailed analysis of photon location to future work.  

\begin{figure}[p]
\includegraphics[width=0.95\linewidth]{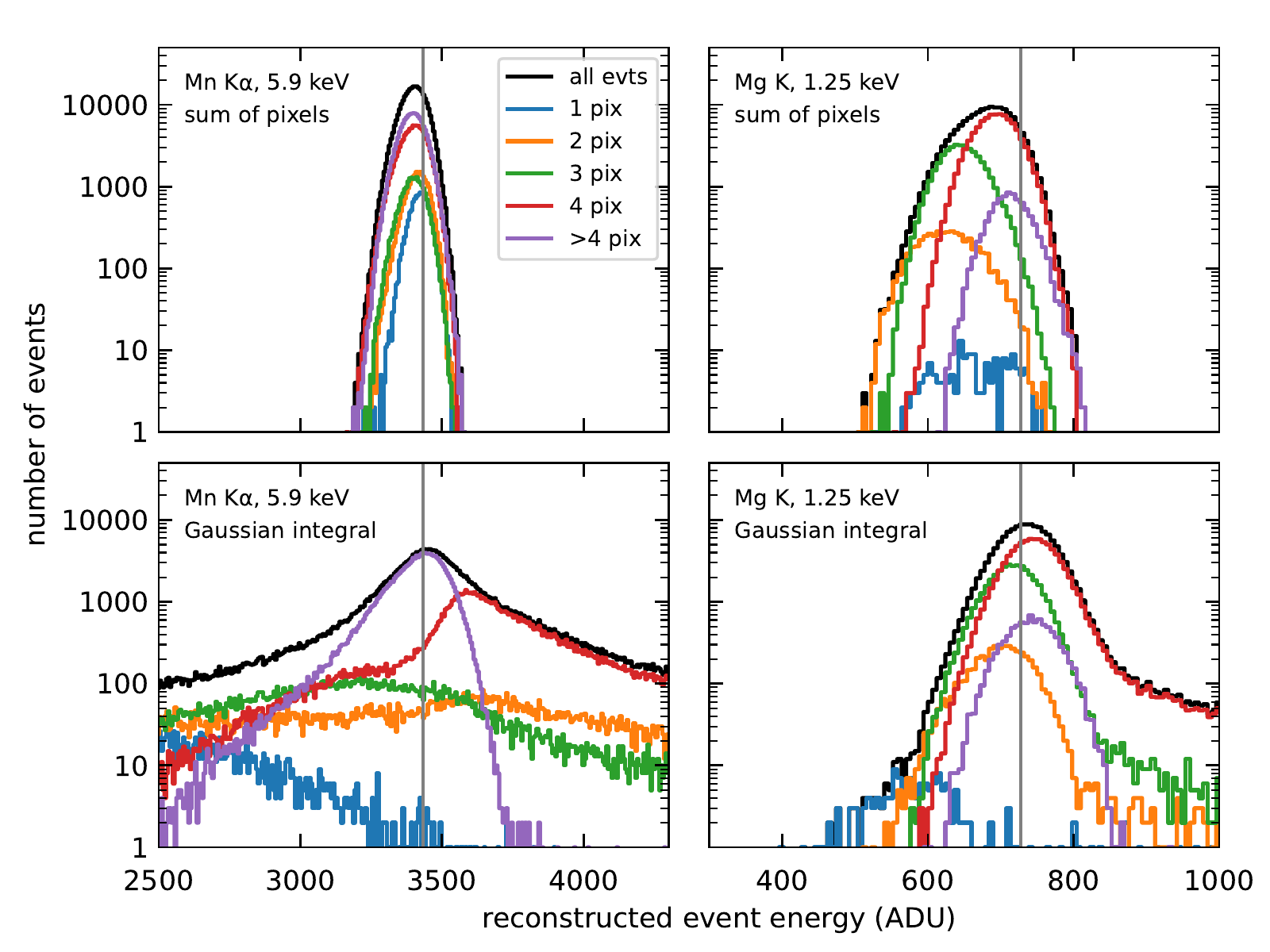}
\caption{Spectra of simulated events from 5.9 keV photons (left panels) and 1.25 keV photons (right panels), with the event energies calculated by summing all pixels above split threshold (top panels) and by integrating under a Gaussian fit (bottom panels). Spectra are separated by the number of pixels above the split threshold, as in Figure \ref{fig:spectra}. At both energies, the Gaussian integral method performs better at estimating the most likely energy of the ensemble of events, as the peak is closer to the expected energy indicated by the vertical line.  The effects of few-pixel events are similar to those shown in Figure \ref{fig:spectra}, and are caused by poor performance of the Gaussian fit for undersampled distributions.}
\label{fig:simspectra}
\end{figure}

These inferences  are generally supported by simulations,  although the details are subtle, as shown in Figure \ref{fig:simspectra}. Here again we separately plot spectra for events with different pixel multiplicities. The results are remarkably similar to  the measured data. At both energies, the Gaussian fitting method does indeed improve the amplitude estimate, the true value of which is known from the simulation inputs: the peaks of the spectral distributions are very close to the input energy. At  5.9 keV, the simulated spectral redistribution is much broader using the Gaussian method than the sum of pixels, just as we observe in the measured spectra.  These broad features are dominated by events with low pixel multiplicities.  At  1.25 keV, the Gaussian fit and pixel summing  estimates produce similar core spectral responses, while the Gaussian fit estimate features an extended high-energy tail populated mainly by four-pixel events. 

In summary, we find that for our devices and at the energies we probed,  although a Gaussian fit to individual events may  provide a slightly less biased amplitude  estimate on average, the effective spectral resolution  is worse at 5.9 keV and comparable at 1.25 keV to that obtained with the traditional sum  of pixels above threshold algorithm.   At the higher energy, the spatial sampling provided by  8 $\mu$m  pixels is generally insufficient to support spatial modelling of individual events. At 1.25 keV, the fitting method recovers some of the sub-threshold signal ignored by the pixel summation algorithm. This results in event amplitude estimates with less bias, but with no less dispersion, than those of the summation method.

\FloatBarrier

\section{Summary and Discussion}
\label{sec:disc}
 
 Our measurements of X-ray-induced charge packets have yielded a number of results. We confirm that charge packets  produced at a given detector depth exhibit (on average)  a Gaussian spatial distribution  to remarkable accuracy. We show that the distribution of charge packet widths,  parameterized by the Gaussian standard deviation $\sigma_{spatial}$, provides useful information  on detector structure. The cumulative width distribution gives the relationship between width and interaction depth. The differential width distribution provides, for a fully depleted detector,  a precise estimate of the size of events produced in the immediate vicinity of the entrance window. Remarkably, this parameter, which is a crucial determinant of detector response to low-energy ($<1$ keV) X-rays, is most easily and accurately measured with higher energy X-rays using the techniques we present here.  The shape of the differential width distribution also provides a clear indication of the extent to which the detector is fully depleted.
 
 We have used  these diagnostics  to tune and validate an implementation  of the   \textsc{Poisson CCD}  simulation. We find  reasonable  agreement with measurements by reducing the amount of (lateral) diffusion, relative to values reported for VRO devices~\cite{Lageetal2021}, by  of order 35\% when the detector is fully depleted.  The poorer agreement for partially depleted configurations indicates that further development development of the simulation is needed.  We will report on this in a future contribution.
 
 We investigated use of Gaussian fits to individual events to estimate their amplitudes. We found that at 5.9 keV this estimator is not as accurate as traditional methods which sum pixel values exceeding a  threshold.  For a significant fraction of  events at this energy,  our detector provides insufficient spatial resolution to measure event shape. At 1.25 keV, the fitting method clearly recovers signal that is neglected by traditional pixel summation algorithms. While this produces a less biased amplitude estimate,  it does not  improve spectral resolution  at the energies we've investigated. This may be a consequence of the relatively low signal-to-noise ratio of the sub-threshold pixels included in the fits.  We speculate that more sophisticated fitting algorithms, for example incorporating priors based on readily measurable event characteristics, may be more successful. We shall also explore this approach in future work. 
 
 Finally, our results highlight an important and perhaps under-appreciated mechanism through which read noise can degrade spectral resolution at lower X-ray energies. It is widely understood that spectral resolution is degraded when charge is shared amongst multiple pixels, since  the  readout noise  associated with each pixel sums in quadrature in the event amplitude calculation. We have shown the  importance of a second mechanism by which readout noise degrades spectral resolution: the effective loss of signal in pixels with values below threshold (see the right-hand panels of Figure~\ref{fig:spectra}). Since the threshold must be a multiple  of readout noise, the magnitude of this lost  sub-threshold signal increases as readout noise increases. In fact, the spectral broadening due to lost, sub-threshold charge in our 1.25 keV data is considerably larger than that due to the noise injected by the sense node amplifier itself, as is demonstrated  in the righthand panels of  Figure \ref{fig:specresponse}. The spectral resolution is much worse than the theoretical (Fano plus readnoise)  expectation, and the charge lost is largest (peak shift 10\% to 20\%)  for pixel multiplicities n=1 and n=2. As a result, the integrated spectral resolution, summed over all events ($\sim 150$ eV FWHM)  is considerably worse  than expected from Fano noise plus the weighted  quadrature sum of readout noise alone ($< 90$ eV FWHM.)

\section{Conclusions and Future Work}
\label{sec:conc}
Mega-pixel X-ray sensors  with large ratios of depletion thickness to pixel size are required for future strategic missions such as  Lynx and AXIS.   We find that direct charge-cloud size measurements  in a \mic{50} thick, \mic{8}-pixel device are useful for validating a basic drift and diffusion simulation of such devices, although more work is required to achieve accurate modeling over a wide range of operating conditions.  Our measurements and simulations suggest that  read-noise dependent,  sub-threshold charge loss may be the most important determinant of  low-energy spectral resolution and it is  therefore essential that this process is fully understood when establishing  sensor noise requirements for these missions. In fact, the detector model described here predicts  that sensors capable of meeting the  low-energy spectral  resolution requirements of AXIS and Lynx  require noise considerably below their notional upper limit  of $4$ electrons RMS. 

To test  this proposition and to make it quantitative, we are extending this work in several ways. We are currently acquiring  data at lower X-ray energies with lower noise detectors. These data will provide more stringent tests of the current simulation.   We are also working to improve the fidelity of the simulation by refining the treatment of lateral diffusion and incorporating a more realistic model of the detector entrance window. We expect this work to lead to more robust detector requirements for future X-ray missions.

\FloatBarrier

\section{Acknowledgements}
We are  deeply saddened by the passing of our co-author and colleague Barry E. Burke. His work advanced imaging technology and enabled the contributions of generations of astronomers to scientific knowledge. He was a most generous, and in the very highest sense, a good human being.

We thank Michelle Gabutti for assistance with data acquisition, Craig Lage for providing key updates to and generous assistance with the \textsc{Poisson CCD} code, and Sven Herrmann for valuable discussions. We gratefully acknowledge support of this  work by NASA through Strategic Astrophysics Technology grants 80NSSC18K0138 and 80NSSC19K0401 to MIT and by MKI's Kavli Research Infrastructure Fund. 
\bibliography{report}   

\begin{thebibliography}{10}

\bibitem{asca}
B.~Burke, R.~Mountain, P.~Daniels, {\em et~al.}, ``{CCD} soft {X}-ray imaging
  spectrometer for the {ASCA} satellite,'' {\em IEEE Transactions on Nuclear
  Science} {\bf 41}, 375--385  (1994).

\bibitem{acis}
G.~P. {Garmire}, M.~W. {Bautz}, P.~G. {Ford}, {\em et~al.}, ``{Advanced CCD
  imaging spectrometer (ACIS) instrument on the Chandra X-ray Observatory},''
  in {\em X-Ray and Gamma-Ray Telescopes and Instruments for Astronomy.},
  J.~E. {Truemper} and H.~D. {Tananbaum}, Eds., {\em Society of Photo-Optical
  Instrumentation Engineers (SPIE) Conference Series} {\bf 4851}, 28--44
  (2003).

\bibitem{xmm}
F.~{Jansen}, D.~{Lumb}, B.~{Altieri}, {\em et~al.}, ``{XMM-Newton observatory.
  I. The spacecraft and operations},'' {\em Astronomy and Astrophysics} {\bf
  365}, L1--L6  (2001).

\bibitem{swift}
N.~{Gehrels}, G.~{Chincarini}, P.~{Giommi}, {\em et~al.}, ``{The Swift
  Gamma-Ray Burst Mission},'' {\em \apj} {\bf 611}, 1005--1020  (2004).

\bibitem{suzaku}
K.~{Mitsuda}, M.~{Bautz}, H.~{Inoue}, {\em et~al.}, ``{The X-Ray Observatory
  Suzaku},'' {\em Publications of the Astronomical Society of Japan} {\bf 59},
  S1--S7  (2007).

\bibitem{erosita}
P.~{Predehl}, R.~{Andritschke}, V.~{Arefiev}, {\em et~al.}, ``{The eROSITA
  X-ray telescope on SRG},'' {\em Astronomy and Astrophysics} {\bf 647}, A1
  (2021).

\bibitem{Gaskin19}
J.~A. {Gaskin}, D.~A. {Swartz}, A.~{Vikhlinin}, {\em et~al.}, ``{Lynx X-Ray
  Observatory: an overview},'' {\em Journal of Astronomical Telescopes,
  Instruments, and Systems} {\bf 5}, 021001  (2019).

\bibitem{AXIS}
R.~{Mushotzky}, ``{AXIS: a probe class next generation high angular resolution
  x-ray imaging satellite},'' in {\em Space Telescopes and Instrumentation
  2018: Ultraviolet to Gamma Ray},  {\em {Proc. SPIE}} {\bf 10699}  (2018).

\bibitem{Groom}
D.~E. {Groom}, S.~E. {Holland}, M.~E. {Levi}, {\em et~al.},
  ``{Back-illuminated, fully-depleted CCD image sensors for use in optical and
  near-IR astronomy},'' {\em Nuclear Instruments and Methods in Physics
  Research A} {\bf 442}, 216--222  (2000).

\bibitem{High2007}
F.~W. {High}, J.~{Rhodes}, R.~{Massey}, {\em et~al.}, ``{Pixelation Effects in
  Weak Lensing},'' {\em \pasp} {\bf 119}, 1295--1307  (2007).

\bibitem{H_marshall}
H.~Marshall, S.~Heine, A.~Garner, {\em et~al.}, ``A small satellite version of
  a soft {x}-ray polarimeter,'' in {\em \procspie},  {\em Space Telescopes and
  Instrumentation: Ultraviolet to Gamma Ray} {\bf 11444}, 11444Y  (2020).

\bibitem{Schattenburg}
M.~Schattenburg, R.~Heilmann, H.~Marshall, {\em et~al.}, ``A diffraction
  limited {W}olter nested-shell telescope concept with pico-radian
  resolution,'' in {\em \procspie},  {\em Space Telescopes and Instrumentation:
  Ultraviolet to Gamma Ray} {\bf 11444}  (2020).

\bibitem{Bautz19}
M.~W. {Bautz}, B.~E. {Burke}, M.~{Cooper}, {\em et~al.}, ``{Toward fast,
  low-noise charge-coupled devices for Lynx},'' {\em Journal of Astronomical
  Telescopes, Instruments, and Systems} {\bf 5}, 021015  (2019).

\bibitem{Falcone2019}
A.~D. {Falcone}, R.~P. {Kraft}, M.~W. {Bautz}, {\em et~al.}, ``{Overview of the
  high-definition x-ray imager instrument on the Lynx x-ray surveyor},'' {\em
  Journal of Astronomical Telescopes, Instruments, and Systems} {\bf 5}, 021019
   (2019).

\bibitem{Hull2019}
S.~V. {Hull}, A.~D. {Falcone}, E.~{Bray}, {\em et~al.}, ``{Hybrid CMOS
  detectors for the Lynx x-ray surveyor high definition x-ray imager},'' {\em
  Journal of Astronomical Telescopes, Instruments, and Systems} {\bf 5}, 021018
   (2019).

\bibitem{Kenter2019}
A.~{Kenter}, R.~{Kraft}, and T.~{Gauron}, ``{Monolithic CMOS detectors for use
  as x-ray imaging spectrometers},'' in {\em UV, X-Ray, and Gamma-Ray Space
  Instrumentation for Astronomy XXI},  {\em Society of Photo-Optical
  Instrumentation Engineers (SPIE) Conference Series} {\bf 11118}, 1111806
  (2019).

\bibitem{Sven}
S.~Herrmann, J.~Wong, T.~Chattopadhay, {\em et~al.}, ``Mcrc v1: development of
  integrated readout electronics for next generation {X}-ray {CCD} detectors
  for future satellite observatories,'' in {\em \procspie},  {\em X-ray,
  Optical,and Infrared Detectors for Astronomy} {\bf 11454}  (2020).

\bibitem{Lageetal2021}
C.~{Lage}, A.~{Bradshaw}, J.~{Anthony Tyson}, {\em et~al.}, ``{Poisson\_CCD: A
  dedicated simulator for modeling CCDs},'' {\em Journal of Applied Physics}
  {\bf 130}, 164502  (2021).

\bibitem{VRO19}
{\v{Z}}.~{Ivezi{\'c}}, S.~M. {Kahn}, J.~A. {Tyson}, {\em et~al.}, ``{LSST: From
  Science Drivers to Reference Design and Anticipated Data Products},'' {\em
  \apj} {\bf 873}, 111  (2019).

\bibitem{Kotov15}
I.~V. {Kotov}, J.~{Haupt}, P.~{Kubanek}, {\em et~al.}, ``{X-ray analysis of
  fully depleted CCDs with small pixel size},'' {\em Nuclear Instruments and
  Methods in Physics Research A} {\bf 787}, 12--19  (2015).

\bibitem{archon}
G.~{Bredthauer}, ``{Archon:} a modern controller for high performance
  astronomical {CCDs},'' in {\em Ground-based and Airborne Instrumentation for
  Astronomy V},  {\em {Proc. SPIE}} {\bf 9147}  (2014).

\bibitem{mpfit}
C.~B. {Markwardt}, ``{Non-linear Least-squares Fitting in IDL with MPFIT},'' in
  {\em Astronomical Data Analysis Software and Systems XVIII},  D.~A.
  {Bohlender}, D.~{Durand}, and P.~{Dowler}, Eds., {\em Astronomical Society of
  the Pacific Conference Series} {\bf 411}, 251  (2009).

\bibitem{Green1990}
M.~A. Green, ``Intrinsic concentration, effective densities of states, and
  effective mass in silicon,'' {\em Journal of Applied Physics} {\bf 67}(6),
  2944--2954  (1990).

\bibitem{Haroetal2020}
M.~S. Haro, G.~Fernandez~Moroni, and J.~Tiffenberg, ``{Studies on Small Charge
  Packet Transport in High-Resistivity Fully Depleted CCDs},'' {\em IEEE
  Transactions on Electron Devices} {\bf 67}(5), 1993--2000  (2020).

\bibitem{Prigozhinetal2021}
I.~Prigozhin, S.~Dominici, and E.~Bellotti, ``{FBMC3D---A Large-Scale 3-D Monte
  Carlo Simulation Tool for Modern Electronic Devices},'' {\em IEEE
  Transactions on Electron Devices} {\bf 68}(1), 279--287  (2021).

\bibitem{GYP2003}
G.~{Prigozhin}, N.~R. {Butler}, S.~E. {Kissel}, {\em et~al.}, ``{An
  experimental study of charge diffusion in the undepleted silicon of {X}-ray
  CCDs},'' {\em IEEE Transactions on Electron Devices} {\bf 50}, 246--253
  (2003).

\bibitem{Miyata02}
E.~{Miyata}, M.~{Miki}, J.~{Hiraga}, {\em et~al.}, ``{Application of the Mesh
  Experiment for the Back-Illuminated Charge-Coupled Device: I. Experiment and
  the Charge Cloud Shape},'' {\em Japanese Journal of Applied Physics} {\bf
  41}, 5827  (2002).

\bibitem{PavlovNousek99}
G.~G. {Pavlov} and J.~A. {Nousek}, ``{Charge diffusion in CCD X-ray
  detectors},'' {\em Nuclear Instruments and Methods in Physics Research A}
  {\bf 428}, 348--366  (1999).

\bibitem{Kotov2021}
I.~V. {Kotov}, S.~{Hall}, D.~{Gopinath}, {\em et~al.}, ``{Analysis of the EMCCD
  point-source response using x-rays},'' {\em Nuclear Instruments and Methods
  in Physics Research A} {\bf 985}, 164706  (2021).

\end{thebibliography}
\bibliographystyle{spiejour}   

\appendix

\section{Tabulation of Spectral Response Parameters}
\label{sec:appendix}

We include here in Tables \ref{tab:specwidths} and \ref{tab:specpeaks} tabulation of the spectral response parameters plotted in Figure \ref{fig:specresponse} for different event pixel multiplicities, as discussed in Section \ref{sec:amp}. The spectra FWHM and peak shift are measured from fitting a single Gaussian to the spectra shown in Figures \ref{fig:spectra} and \ref{fig:simspectra}.

\begin{table}[p]
\footnotesize
\begin{center}
\caption{Spectral resolution using different event reconstruction methods.} \label{tab:specwidths}
\begin{tabular}{lcccccc}\hline \hline
        &        & Fano    &\multicolumn{2}{c}{Measurements$^b$} & \multicolumn{2}{c}{Simulations$^b$} \\
        &        & + noise$^a$ & Pixel sum & Gauss.~int. & Pixel sum & Gauss.~int. \\
Energy  & \#~pix & FWHM    & FWHM & FWHM & FWHM & FWHM \\
        &        & (eV)    & (eV) & (eV) & (eV) & (eV) \\
\hline
5.9 keV  & any &  ... &  217 & ~~601 &  182 & ~~545 \\
         & 1   &  123 &  144 &  ~~~~~3 &  138 &  1120 \\
         & 2   &  129 &  173 &  6030 &  157 &  4044 \\
         & 3   &  135 &  204 &  1784 &  168 &  1701 \\
         & 4   &  140 &  212 & ~~608 &  192 & ~~635 \\
         & 5   &  146 &  226 & ~~444 &  180 & ~~427 \\
         & 6   &  151 &  217 & ~~306 &  174 & ~~286 \\
         & 7   &  156 &  216 & ~~298 &  165 & ~~279 \\
         & 8   &  160 &  259 & ~~304 &  169 & ~~256 \\
\hline
1.25 keV & any &  ... &  179 & ~~174 &  157 & ~~151 \\
         & 1   & ~~66 &  257 & ~~758 &  233 & ~~218 \\
         & 2   & ~~77 &  149 & ~~171 &  176 & ~~163 \\
         & 3   & ~~86 &  126 & ~~154 &  123 & ~~137 \\
         & 4   & ~~94 &  126 & ~~152 &  120 & ~~140 \\
         & 5   &  102 &  134 & ~~162 &  104 & ~~140 \\
         & 6   &  109 &  ... & ~~... &  106 & ~~128 \\
\hline
\multicolumn{7}{p{.70\textwidth}}{$a$: `Fano + noise' is the theoretical Fano-limit spectral FWHM added in quadrature with Gaussian readout noise in each pixel.} \\
\multicolumn{7}{p{.70\textwidth}}{$b$: Conversion  from measured analog-to-digital units (ADU) to eV uses a gain factor derived from the peak of the 5.9 keV  histogram of  events with pixel multiplicity $\ge 5$\,  (see Fig, \ref{fig:spectra}a).The simulations assume an electron liberation energy of 3.65 eV  to convert to energy, and the same measured gain factor of 1.718 eV ADU$^{-1}$ to compare to lab data ADU values. } \\
\end{tabular}
\end{center}
\normalsize
\end{table}

\begin{table}[p]
\footnotesize
\begin{center}
\caption{Spectral peak shift$^a$ using different event reconstruction methods.}\label{tab:specpeaks}
\begin{tabular}{lccccccc}\hline \hline
        &        & \multicolumn{3}{c}{Measurements}                   & \multicolumn{3}{c}{Simulations} \\
        &        & frac.  & Pixel sum & Gauss.~int. & frac.  & Pixel sum & Gauss.~int. \\
Energy  & \#~pix & events & peak shift    & peak shift        & events & peak shift    & peak shift \\
\hline
5.9 keV  & any & 1    &  $-$1.2\% & ~~$+$1.0\% & 1     &  $-$0.9\% & ~~$+$0.6\% \\
         & 1   & 0.02 &  $-$0.7\% &  $-$39.4\% & 0.04  &  $-$0.3\% &  $-$44.3\% \\
         & 2   & 0.06 &  $-$0.8\% &  $-$13.1\% & 0.08  &  $-$0.5\% & ~~$+$1.0\% \\
         & 3   & 0.06 &  $-$1.1\% & ~~$-$6.8\% & 0.07  &  $-$1.0\% & ~~$-$7.6\% \\
         & 4   & 0.30 &  $-$1.3\% & ~~$+$0.9\% & 0.35  &  $-$0.9\% & ~~$+$5.9\% \\
         & 5   & 0.16 &  $-$1.6\% & ~~$+$1.4\% & 0.14  &  $-$1.5\% & ~~$+$1.0\% \\
         & 6   & 0.25 &  $-$1.2\% & ~~$+$0.1\% & 0.22  &  $-$1.0\% & ~~$+$0.2\% \\
         & 7   & 0.13 &  $-$0.7\% & ~~$-$0.9\% & 0.09  &  $-$0.6\% & ~~$-$1.2\% \\
         & 8   & 0.02 &  $+$0.0\% & ~~$-$1.4\% & 0.004 &  $+$0.0\% & ~~$-$2.3\% \\
\hline
1.25 keV & any & 1    & ~~$-$7.6\% & ~~$-$1.1\% & 1     & ~~$-$6.0\% & ~~$+$1.1\% \\
         & 1   & 0.002 &  $-$13.2\% &  $-$49.8\% & 0.001 & ~~$-$8.6\% &  $-$20.3\% \\
         & 2   & 0.082 &  $-$16.5\% & ~~$-$5.9\% & 0.03  &  $-$14.0\% & ~~$-$3.4\% \\
         & 3   & 0.353 &  $-$11.3\% & ~~$-$3.3\% & 0.27  &  $-$11.3\% & ~~$-$1.5\% \\
         & 4   & 0.526 & ~~$-$4.9\% & ~~$+$1.2\% & 0.63  & ~~$-$4.6\% & ~~$+$2.4\% \\
         & 5   & 0.034 & ~~$-$0.4\% & ~~$+$0.7\% & 0.06  & ~~$-$2.2\% & ~~$+$1.9\% \\
         & 6   & 0.000 & ~~...      & ~~...      & 0.002 & ~~$+$2.6\% & ~~$+$4.9\% \\
\hline
\multicolumn{8}{p{.77\textwidth}}{$a$: `Peak shift' is defined as the difference between the emission line energy and the peak (mode) of a Gaussian fit to the spectrum.} \\
\end{tabular}
\end{center}
\normalsize
\end{table}

\end{spacing}
\end{document}